# An Inverse-Designed Multilayer Selective Emitter for High-Efficiency Thermophotovoltaics

Md Anwar Hosen[1], Kavyashree Keremane[2], Bed Poudel[2], Linxiao Zhu[1]*

[1]*Department of Mechanical Engineering, The Pennsylvania State University, University Park, Pennsylvania 16802, USA*

[2] *Department of Materials Science and Engineering, The Pennsylvania State University, University Park, Pennsylvania 16802, USA*

**Contact author: lqz5242@psu.edu*

## ABSTRACT

High thermophotovoltaic (TPV) conversion efficiencies have typically relied on photovoltaic cells that reflect sub-bandgap radiation back toward the emitter. Here, we numerically investigate an alternative approach in which spectral selectivity is engineered primarily on the emitter side. We optimize an emitter consisting of 16-layer aperiodic MgO/Si coating on tungsten, for coupling to a GaSb cell with an antireflection coating and an Au mirror. A detailed-balance framework incorporates polarization- and angle-resolved emissivity, multiple reflections, and Auger nonradiative recombination. At an emitter temperature of 1500 K, the optimized structure suppresses 97.1% of the incident-blackbody-equivalent sub-bandgap power at the source; a further 1.9% is returned to the emitter by the cell, leaving 1.0% absorbed parasitically in the cell. The resulting modeled radiative-to-electrical conversion efficiency reaches 47.0%, compared with 30.8% for a SiC emitter paired with the same cell, while maintaining a comparable maximum electrical power density. These findings indicate that emitter-side spectral engineering could substantially relax cell-side photon-recycling requirements and provide significant improvement in TPV conversion.

## I. INTRODUCTION

While solar cells produce electricity from the thermal radiation of the sun, thermophotovoltaic (TPV) systems generate electricity from locally supplied heat by coupling a hot emitter to a photovoltaic (PV) cell [1,2]. Because the heat can come from combustion, concentrated sunlight, industrial waste heat, stored thermal energy, or nuclear reaction, TPV is a promising route for distributed power generation [1,3], waste-heat recovery [4,5], solar energy harvesting [6,7] and high-temperature thermal energy storage[8,9]. For such systems to become practical and cost-effective, both the conversion efficiency and the power density need to be improved [2,10,11].

To achieve high-performance TPV operation, it is crucial to have selective radiative heat transfer. The cell must absorb high-energy (in-band) photons efficiently, while low-energy (out-of-band, OOB) photons should either not be emitted by a selective emitter or be reflected by the cell back to the emitter, where they are reabsorbed and recycled [2,12]. Any OOB photon absorbed in the cell causes parasitic heating of the cell, in turn raising the requirement of thermal management for the cell.

Recent demonstrations have greatly improved TPV performance by engineering selective cells. For cells with Au rear mirrors with sub-bandgap reflectance of 93–95%, radiative-to-electrical conversion efficiency has reached 29.1% at 1207 °C [12], 38.8% at 1850 °C [13], and 41.1% using multijunction III–V cells at 2400 °C [14]. Air-bridge cells, where the absorber is suspended above an Au mirror by a submicron air gap, raised the OOB reflectance to about 98.5% and the efficiency to 32% at 1182 °C [15] and 43.8% at 1435 °C [16]. A 5% OOB reflectance loss lowers the efficiency by about 10%[2,15], and raising the OOB reflectance from 95.3% to 98.5% increased the measured efficiency from 23.4% to 32.0%[15]. However, having high OOB reflectance has relied on complex architectures including air-bridge cells[15], or multijunction cells[14] facing emitter at 2400 °C that may pose material-stability challenges. Moreover, in all these demonstrations the emitter is a nearly gray body such as SiC, graphite or a tungsten-filament lamp [12–16], so rejecting OOB radiation almost entirely relies on the cell.

On the emitter side, spectrally selective emitters[17,18] have been developed using photonic crystals [19,20], metamaterials and metasurfaces [21–24], and multilayer coatings [25–32]. Several of them were obtained by inverse design using Bayesian optimization, normalizing flows, and reinforcement learning [24,26,29,32]. Multilayer emitters on tungsten (W) are particularly attractive because they are planar and can be fabricated by standard thin-film deposition without lithography [17,18]. Recent design of multilayer emitters have shown stability up to 1423–1473 K [28,30,31]. In particular, $SiO_2$/Si multilayers on W have been fabricated at wafer scale with an emissivity below 0.01 between 1.7 and 7.5 μm apart from a few narrow peaks, although their spectral efficiency was evaluated only up to 8 μm, which excludes the phonon polariton resonances of $SiO_2$, and non-radiative recombination was omitted[30]. Nevertheless, most demonstrated emitters still have OOB emissivities of 0.15–0.3 [17]. Measured system efficiencies with selective emitters, including solar

TPV, remain 7.5–24% [20,21,33], and predicted efficiencies with GaSb cells range from 27% to 41% [28,30].

Moreover, many studies of selective emitters evaluate performance with emissivity at normal direction, an idealized cell, or a limited spectral range [22,24,26,29,30,34,35], and some neglect photon recycling [32] or nonradiative recombination [28, 29, 36]. For example, efficiencies of 47–59% have been predicted for single and multilayer coatings on W with GaSb cells, using an ideal cell and spectra limited to 3 μm[36] or shown only up to 5 μm[28], whereas about a quarter of the blackbody power at 1000 °C lies beyond 5 μm. Since neglecting nonradiative recombination overestimates the efficiency [36], such models may overestimate the performance.

It has also been argued that, once the rear reflectivity of the cell is high, spectral control of the emitter becomes unnecessary [12]. A techno-economic analysis found that emitter-side control competes with cell-side control only when the sub-bandgap emissivity falls to about 0.1 or below [11]. In a planar system, however, the emitter and the cell exchange radiation through multiple reflections, so the OOB loss depends on both the emitter and the cell[15,16], and multilayer emitters can now reach sub-bandgap emissivities well below 0.1 [30]. It would be therefore interesting to investigate whether a selective emitter when combined with a conventional cell can reach an OOB loss in the cell as low, and an efficiency as high, as those of the best cell-engineered systems[15,16] when angle-dependent emission and nonradiative recombination are taken into account.

In this article, we consider a highly selective multilayer emitter combined with a conventional GaSb cell that has a standard antireflection coating and an Au rear mirror. The emitter consists of 16 alternating MgO and Si layers on a W substrate which is optimized for the 0.725 eV bandgap of GaSb. The cell is not optimized in this work, so that the improvement can be attributed to the emitter. We show that, because the emitter suppresses most of the sub-bandgap emission and the cell also reflects back to the emitter, about 99% of the sub-bandgap blackbody power is kept out of the cell, comparable to the best air-bridge cells[15,16]. We describe the structures and the model in Sec. II, present the results in comparison with a SiC emitter in Sec. III, discuss them in Sec. IV, and conclude in Sec. V.

## II. SYSTEM AND MODEL

We consider a planar TPV system in which a hot selective emitter faces a GaSb cell, as schematically shown in Figure 1(a). The hot emitter radiates photons toward the PV cell. In-band photons are absorbed by the cell, while OOB photons that reach the cell are reflected by the rear mirror back to the emitter. The emitter, shown in Figure 1(b), consists of 16 alternating layers of MgO and Si with a total dielectric thickness of 5.99 μm on W substrate. The layer facing the cell is MgO, and the layer adjacent to W is Si. The layer thicknesses are aperiodic and obtained by inverse design of the TPV efficiency. While MgO and W were chosen for high-temperature use [18,37], Si limits the maximum operating temperature, as discussed in Sec. IV. W is a refractory metal and has a high infrared reflectance, which suppresses OOB emission. MgO and Si provide a large refractive-index contrast in the near infrared. MgO has a melting point of 2852 °C [18,38], and all-oxide multilayers containing MgO have remained structurally stable during heating to 1000 °C in air [38].

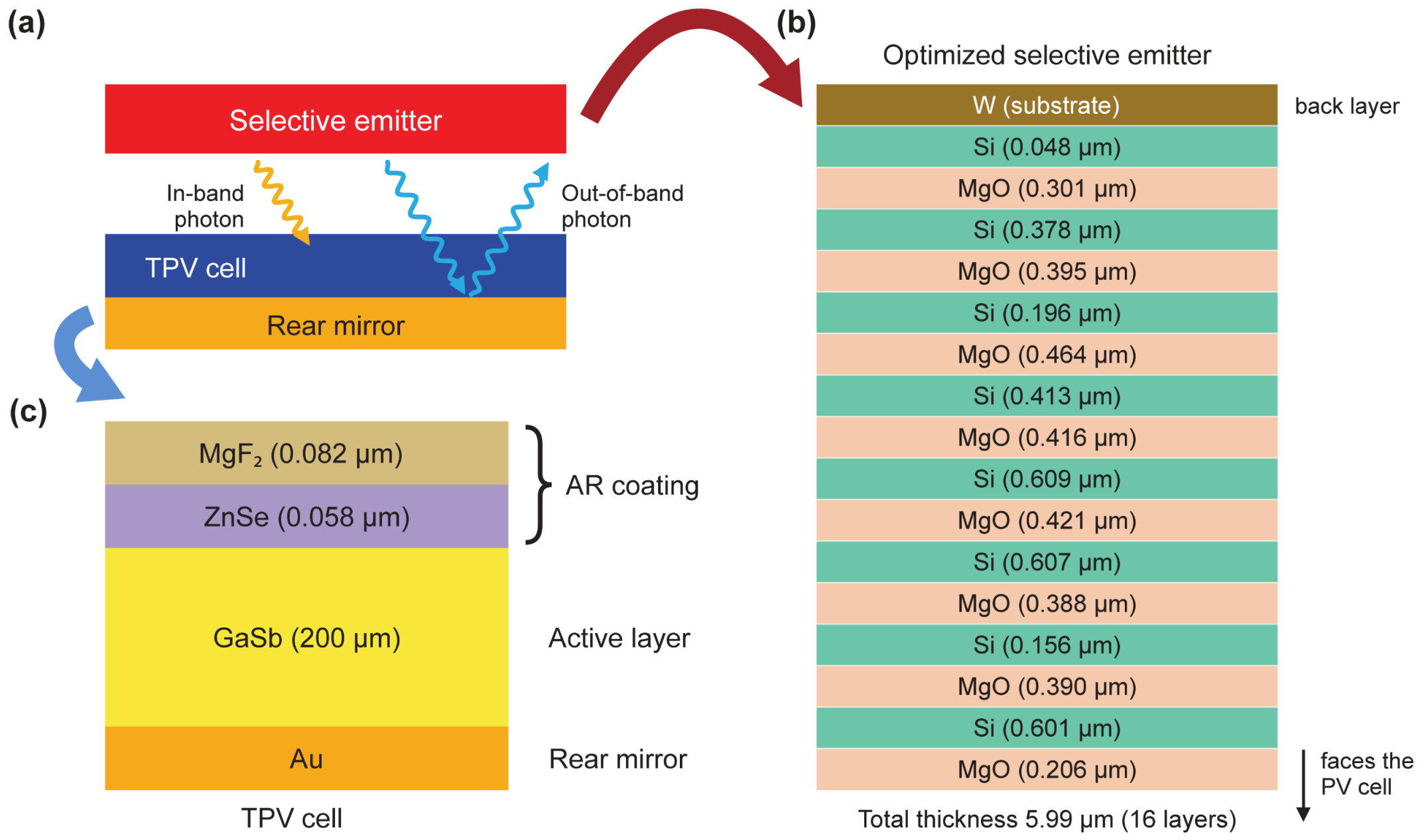

**Figure 1:** Thermophotovoltaic system with a multilayer selective emitter and a GaSb cell. (a) Schematic of the planar TPV system. The emitter radiates photons towards the cell . In-band photons (yellow) are absorbed in the cell, and out-of-band photons (blue) are reflected by the rear mirror back to the emitter. (b) Optimized selective emitter, 16 alternating MgO and Si layers with the labeled thicknesses on W substrate. The bottom MgO layer faces the cell. (c) GaSb cell consisting of an $MgF_2$/ZnSe antireflection (AR) coating, a 200 μm GaSb active layer, and an Au rear mirror.

The cell, shown in Figure 1(c), is a conventional GaSb cell. It consists of a 200 μm GaSb active layer with a bandgap of $E_g = 0.725$ eV, corresponding to a cutoff wavelength of 1.71 μm, a standard two-layer antireflection coating of $MgF_2$ and ZnSe on the front, and an Au rear mirror. Its OOB reflectance of about 93% is typical of Au-backed cells [2,12,14] and lower than that of air-bridge cells [15,16]. As a reference, we consider a bulk SiC emitter, which is widely used in TPV experiments [15,16]. Its emissivity, calculated from the optical constants of SiC[39] for a thick flat surface, is nearly gray, about 0.7–0.75 in the spectral range of interest.

We model the spectral, directional emissivity of the emitter $\varepsilon_{e,p}(\omega,\theta)$ and the absorptivity of the cell $\varepsilon_{c,p}(\omega,\theta)$ for transverse-electric (TE) and transverse-magnetic (TM) polarizations ($\mathrm{p} = \mathrm{TE,TM}$), using the scattering-matrix formalism.[40] From Kirchhoff's law, the emissivity of the emitter equals its absorptivity $\varepsilon_{e,p} = 1 - R_{e,p}$ for the opaque emitter. The calculations cover wavelengths from 0.5 to 20 μm and polar angles from 0° to 89.5° in steps of about 1°. The optical constants of GaSb are taken from Ref. [41] and Ref. [42]. The optical constants of the other materials are taken from Ref. [41] and Ref. [39].

Because the emitter and the cell are planar and parallel, radiation can undergo multiple reflections between them. The effective emissivity of the emitter–cell pair, which accounts for the multiple reflections and which we evaluate separately for each polarization $p$, is [15,16]:

$$\varepsilon_{\mathrm{eff},p}(\omega,\theta) = \frac{\varepsilon_{e,p}(\omega,\theta)\varepsilon_{c,p}(\omega,\theta)}{\varepsilon_{e,p}(\omega,\theta) + \varepsilon_{c,p}(\omega,\theta) - \varepsilon_{e,p}(\omega,\theta)\varepsilon_{c,p}(\omega,\theta)}. \quad (1)$$

The polarization-averaged effective emissivity is $\varepsilon_{\mathrm{eff}}(\omega,\theta) = \left[\varepsilon_{\mathrm{eff,TE}}(\omega,\theta) + \varepsilon_{\mathrm{eff,TM}}(\omega,\theta)\right]/2$. The radiative current density in the cell at voltage bias $V$ is [43]

$$J_{\mathrm{rad}}(V) = \frac{q}{2\pi^2 c^2}\int_{\omega_g}^{\infty} d\omega \int_0^{\frac{\pi}{2}} d\theta \cos\theta \sin\theta\, \varepsilon_{\mathrm{eff}}(\omega,\theta)\, \omega^2 [f_{\mathrm{BE}}(\omega,T_e,0) - f_{\mathrm{BE}}(\omega,T_c,qV)], \quad (2)$$

where $q$ is the charge of a proton, $c$ is the speed of light in vacuum, $T_e$ and $T_c$ are the temperatures of the emitter and the cell, $\omega_g = E_g/\hbar$ is the bandgap angular frequency, $f_{\mathrm{BE}}(\omega,T,\mu) = 1/[\exp[(\hbar\omega - \mu)/k_B T] - 1]$ is the Bose–Einstein distribution, $\hbar$ is the reduced Planck constant, and $k_B$ the Boltzmann constant. For non-radiative recombination, we consider the Auger recombination current density

$$J_{\mathrm{Auger}}(V) = q t_c\left[C_n n(V) + C_p p(V)\right]\left[n(V)p(V) - n_i^2\right], \quad (3)$$

where $t_c = 200$ μm is the cell thickness, $C_n$ and $C_p$ are the Auger coefficients for electrons and holes, respectively, $n(V)$ and $p(V)$ are the electron and hole concentrations, respectively, and $n_i$ is the intrinsic carrier concentration. For an undoped cell, $n = p = n_i \exp(qV/2k_B T_c)$. We use $C_0 = C_p + C_n = 5 \times 10^{-30}$ $\mathrm{cm^6\ s^{-1}}$ and $n_i = 1.5 \times 10^{12}$ $\mathrm{cm^{-3}}$ for GaSb cell[44,45]. The net current density is $J(V) = J_{\mathrm{rad}}(V) - J_{\mathrm{Auger}}(V)$, and the output power density is $P_{\mathrm{out}}(V) = VJ(V)$.

The radiative power absorbed by the cell is

$$P_{\mathrm{in}} = \frac{1}{2\pi^2 c^2}\int_0^{\infty} d\,\omega \int_0^{\frac{\pi}{2}} d\,\theta\, \cos\theta \sin\theta\, \varepsilon_{\mathrm{eff}}(\omega,\theta)\, \hbar\omega^3 f_{\mathrm{BE}}(\omega, T_e, 0), \tag{4}$$

and the conversion efficiency is

$$\eta = \frac{\max\limits_{V}[VJ(V)]}{P_{\mathrm{in}}}. \tag{5}$$

Out-of-band photons that are not absorbed by the cell return to the emitter and are not counted in $P_{\mathrm{in}}$. To separate the spectral and electrical losses, we evaluate the spectral efficiency [2,15]

$$\eta_{\mathrm{SE}} = \frac{\omega_g \int_{\omega_g}^{\infty} \bar{\varepsilon}_{\mathrm{eff}}(\omega)\, \omega^2 f_{\mathrm{BE}}(\omega, T_e, 0)\, d\omega}{\int_0^{\infty} \bar{\varepsilon}_{\mathrm{eff}}(\omega)\; \omega^3 f_{\mathrm{BE}}(\omega, T_e, 0)\, d\omega}, \tag{6}$$

$\eta_{SE}$ is the fraction of the power absorbed in the cell that would be converted to electricity if every absorbed above-bandgap photon delivered exactly the bandgap energy $\hbar\omega_g$. It is reduced by sub-bandgap absorption in the cell and by thermalization of above-bandgap photons, and through $\bar{\varepsilon}_{\mathrm{eff}}(\omega)$, which includes the multiple reflections between the emitter and the cell. Here, $\bar{\varepsilon}_{\mathrm{eff}}(\omega)$ is the hemispherical effective emissivity of the emitter–cell pair,

$$\bar{\varepsilon}_{\mathrm{eff}(\omega)} = \frac{\int_0^{\frac{\pi}{2}} \varepsilon_{\mathrm{eff}}(\omega,\theta) \cos\theta \sin\theta\, d\theta}{\int_0^{\frac{\pi}{2}} \cos\theta \sin\theta\, d\theta}. \tag{7}$$

We also report the fraction of the absorbed power that lies below the bandgap, $\frac{P_{below}}{P_{in}}$, where $P_{below}$ is given by Eq. (4) with the upper limit of the frequency integral replaced by $\omega_g$.Throughout this work, the cell temperature is $T_c = 300$ K.

## III. RESULTS

Figure 2(a) shows the normal-incidence emissivity of the optimized emitter together with the absorptivity of the GaSb cell. The emitter shows a high-emissivity band that rises sharply near the bandgap and extends to about 1.05 eV, with an average emissivity of 0.88 between 0.725 and 1.05 eV. This band covers the in-band part of the blackbody spectrum at typical TPV temperatures, as shown by the 1300 K blackbody spectrum. Below the bandgap, the emissivity is close to zero except for a few narrow resonances, which arise from absorption in W at the resonances of the stack and from the phonon absorption of MgO at long wavelengths. Because these resonances are narrow, the blackbody-weighted emissivity below the bandgap at 1300 K is only 0.033 at normal incidence and 0.030 for hemispherical emissivity. Above 1.05 eV, the emissivity varies strongly with photon energy, but this range carries a small fraction of the blackbody power at the temperatures considered here. The cell absorptivity rises at the bandgap and reaches about 0.77 at 0.8 eV and nearly unity at 1.6 eV. Below the bandgap, the cell absorbs on average about 7% of the incident blackbody radiation, mainly through parasitic absorption in the Au mirror, which corresponds to an OOB reflectance of 93% at 1300 K, close to the values reported for Au-backed cells [2,12,14].

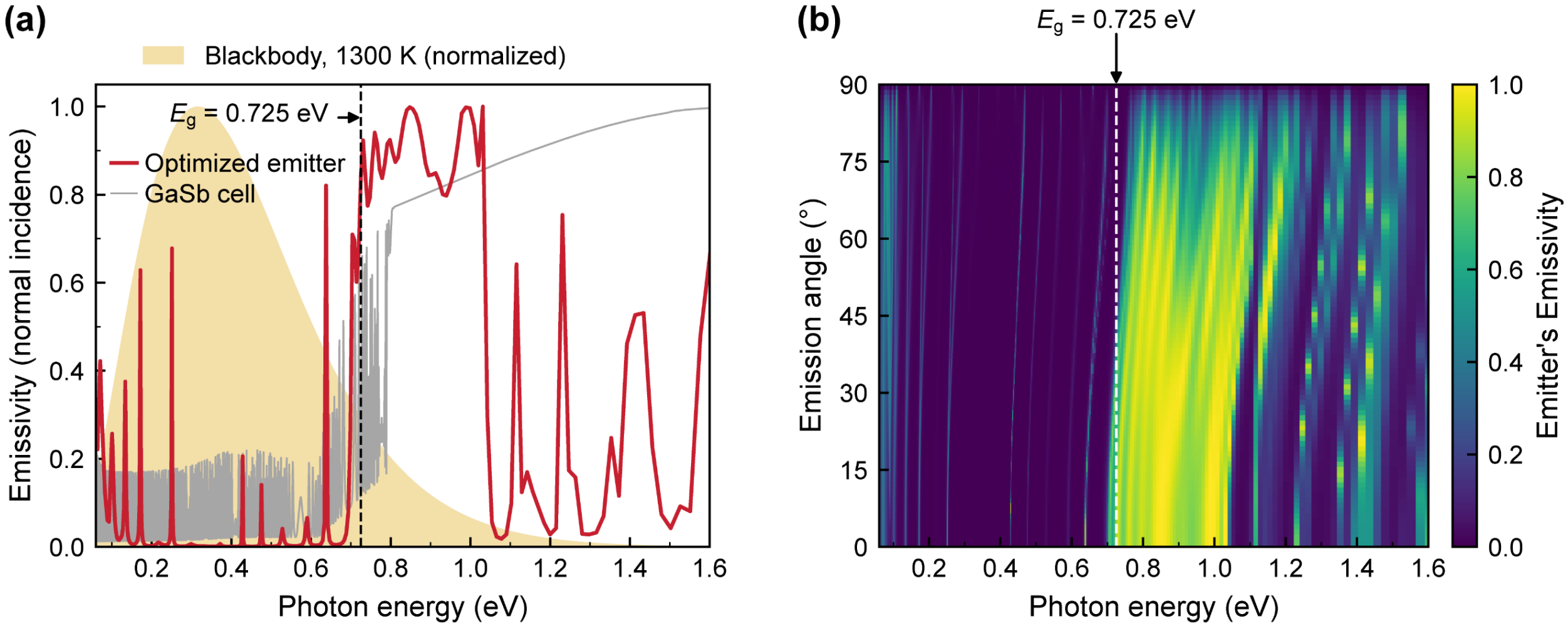

**Figure 2.** Spectral properties of the selective emitter and the GaSb cell. (a) Emissivity of the optimized emitter (red) and absorptivity of the GaSb cell (gray) at normal incidence versus photon energy, averaged over the two polarizations. The shaded area is the normalized blackbody spectrum at 1300 K. The vertical dashed line marks the GaSb bandgap $E_g = 0.725$ eV. (b) Emissivity of the optimized emitter versus photon energy and angle, averaged over the two polarizations. The white dashed line marks $E_g$.

Figure 2(b) shows the emissivity of the emitter as a function of photon energy and emission angle. The high-emissivity band persists up to an emission angle of about 60°, which covers most of the hemispherical weight $\cos\theta\sin\theta$ that peaks at 45°. The average emissivity between the bandgap and 1.05 eV is 0.88, 0.90, 0.85 and 0.72 at 0°, 30°, 45° and 60°, respectively, and decreases to 0.62 at 70°. The lower edge of the band shifts from 0.70 eV at normal incidence to 0.74 eV at 45° and 0.76 eV at 60°. The sub-bandgap resonances also shift with angle but remain narrow at all angles. The blackbody-weighted hemispherical emissivity at 1300 K is 0.72 in band and 0.030 below the bandgap. We note that the hemispherical in-band emissivity of the selective emitter is almost the same as that of SiC (0.70 at 1500 K for both), so the selective emitter does not sacrifice in-band emission.

To show where the sub-bandgap radiation goes, we divide the hemispherical blackbody spectrum B(ω) into four parts, as shown in Figures 3(a) and 3(b) for an emitter temperature of 1500 K. (i) The part $(1-\bar{\varepsilon}_e)B$ is suppressed: it is never emitted by the emitter, where $\bar{\varepsilon}_e$ is the hemispherical emissivity of the emitter. (ii) The part $(\bar{\varepsilon}_e-\bar{\varepsilon}_{\mathrm{eff}})B$ is recycled: it is emitted, but the cell reflects it back to the emitter, where it is reabsorbed, so this energy is not lost. (iii) The part $\bar{\varepsilon}_{\mathrm{eff}}B$ below the bandgap is absorbed in the cell but generates no current, so it contributes to parasitic heating. (iv) The part $\bar{\varepsilon}_{\mathrm{eff}}B$ above the bandgap is absorbed in the cell and is the useful power. Therefore, only part (iii) is a loss below the bandgap.

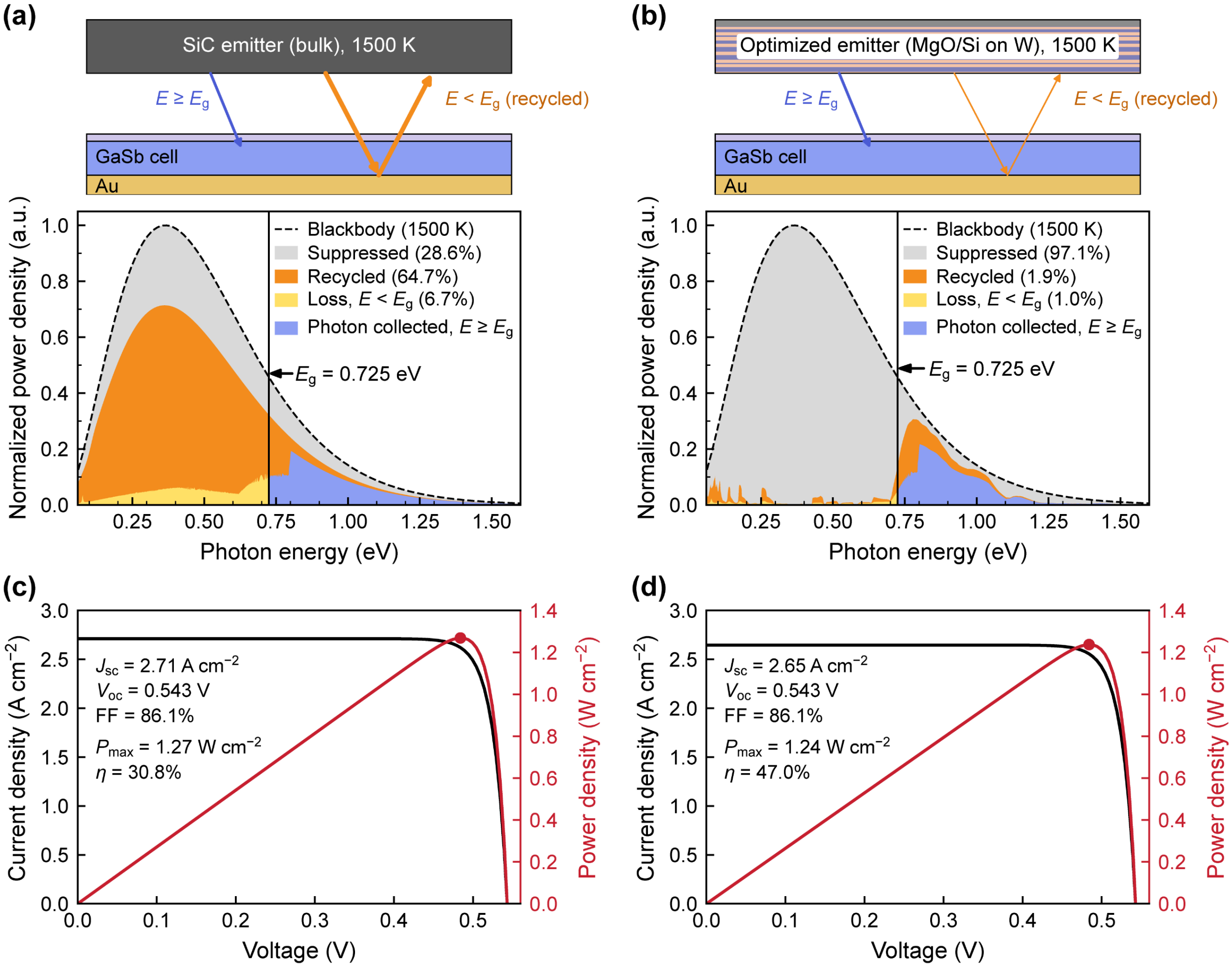

**Figure 3.** Power partition and current–voltage relation at an emitter temperature of 1500 K. (a), (b) Hemispherical blackbody spectral power at 1500 K (dashed), normalized to its maximum, partitioned for (a) the SiC emitter and (b) the optimized emitter, each facing the GaSb cell, as illustrated by the schematics above the panels. Gray: suppressed, $(1-\bar{\varepsilon}_e)B$; orange: recycled, $(\bar{\varepsilon}_e-\bar{\varepsilon}_{eff})B$; yellow: absorbed in the cell below the bandgap (loss), $\bar{\varepsilon}_{eff}B$ for $E < E_g$; blue: absorbed photon in the cell above the bandgap (useful), $\bar{\varepsilon}_{eff}B$ for $E > E_g$. The percentages noted here are fractions of the sub-bandgap blackbody power. For each emitter, the suppressed, recycled, and absorbed fractions add up to 100%. The vertical line marks $E_g$. (c), (d) Current density (black, left axis) and output power density (red, right axis) versus voltage for (c) the SiC emitter and (d) the optimized emitter. The dots mark the maximum power point. The cell is maintained at 300 K.

The sub-bandgap blackbody power is split into three parts that add up to 100%, suppressed, recycled, and absorbed in the cell. Only the last part is a loss. For the SiC emitter [Figure 3(a)], which is nearly black, 28.6% is suppressed, 64.7% is recycled, and the remaining 6.7% is absorbed in the cell as a loss. For the optimized emitter [Figure 3(b)], 97.1% is suppressed, 1.9% is recycled, and only 1.0% is absorbed in the cell, 6.7 times less than for SiC. Thus, with the same conventional cell, the fraction of the sub-bandgap power that does not reach the cell (suppressed plus recycled) increases from 93.3% to 99.0%. This is comparable to the 97.2–98.5% sub-bandgap reflectance of air-bridge cells[15,16], but in our case it comes mainly from the emitter rather than from the cell. Suppression can also be more robust than recycling, as recycling works only if the reflected photons return to the emitter and becomes less effective with a finite view factor between the emitter and the cell. Above the bandgap, the power absorbed in the cell is similar for the two emitters, since their hemispherical in-band emissivities are similar.

Figures 3(c) and 3(d) show the current density and output power density versus voltage at 1500 K. The two emitters give similar short-circuit current densities, $J_{sc} = 2.71$ A cm$^{-2}$ for SiC and 2.65 A cm$^{-2}$ for the optimized emitter, and the same open-circuit voltage of 0.543 V and fill factor of 86.1%. The maximum output power densities are also similar, 1.27 and 1.24 W cm$^{-2}$, respectively. However, because the optimized emitter delivers far less sub-bandgap power to the cell, the conversion efficiency increases from 30.8% for SiC to 47.0%. This shows that the efficiency gain comes entirely from the reduction of the absorbed heat, without a penalty in output power.

Figure 4 compares the performance of the two emitters for emitter temperatures from 900 to 1700 K. The efficiency of the optimized emitter is 17.7% at 900 K, 34.7% at 1100 K, 43.1% at 1300 K, and 47.0% at 1500 K [Figure 4(a)]. For SiC emitter, the efficiency is 5.0%, 14.0%, 23.5% and 30.8% at 900 K, 1100 K, 1300 K, and 1500 K, respectively. The improvement is largest at intermediate temperatures, reaching 20.8% absolute increase at 1200 K, where the blackbody spectrum still lies mostly below the bandgap and the suppression of sub-bandgap emission matters most. At higher temperatures, a larger fraction of the blackbody spectrum lies above the bandgap, so the efficiency of SiC increases faster, and the gain decreases to 16.2% at 1500 K. The selective emitter achieves at 1100 K roughly the efficiency that SiC achieves at 1700 K, which allows the system to operate at a much lower temperature for the same efficiency.

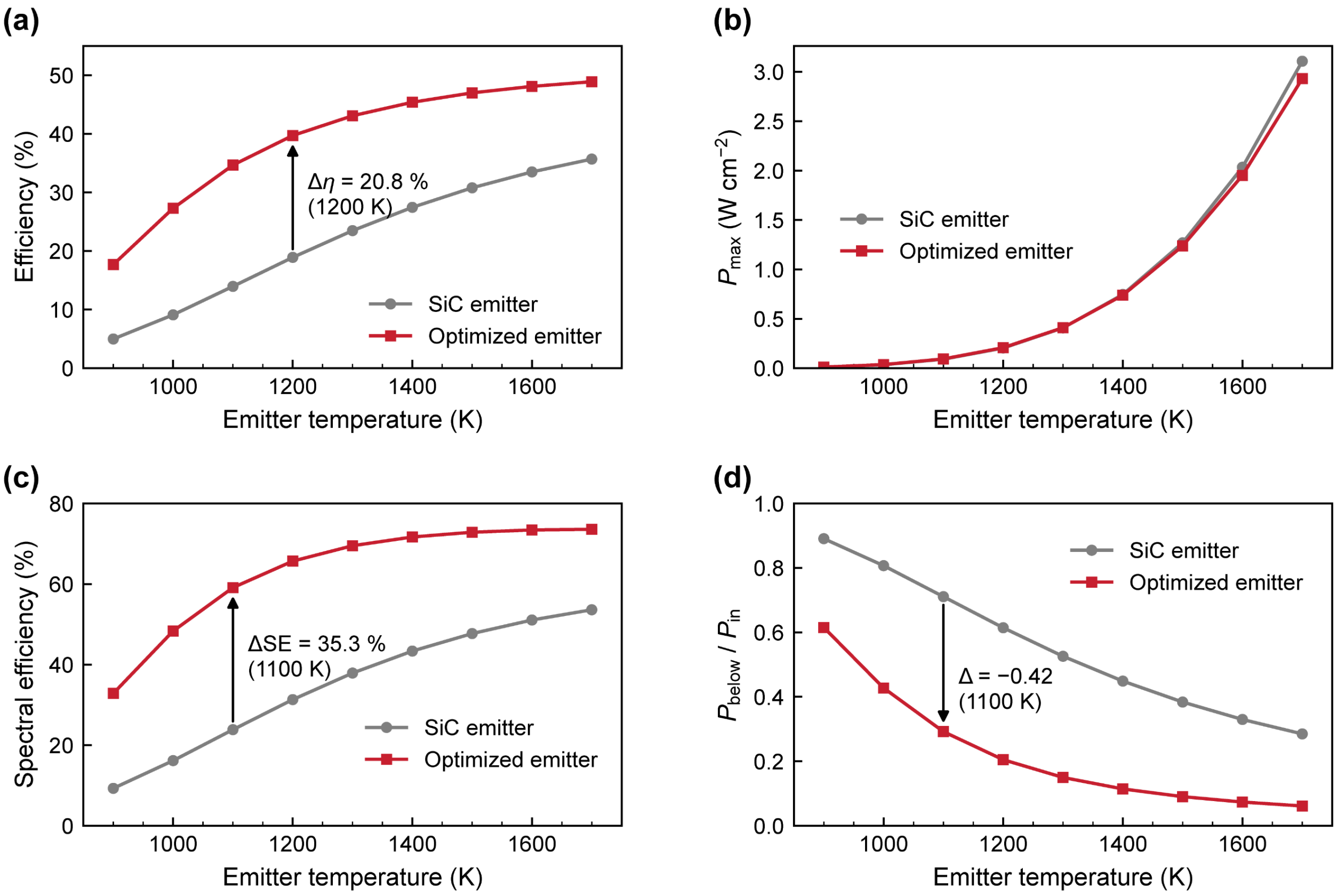

**Figure 4.** Performance of the GaSb TPV system with the SiC emitter (gray circles) and the optimized emitter (red squares). (a) Conversion efficiency. (b) Maximum output power density. (c) Spectral efficiency. (d) Fraction of the absorbed power that lies below the bandgap, $P_{below}/P_{in}$. The arrows mark the temperatures with the largest difference between the two emitters. The emitter temperature is varied from 900 K to 1700 K, with the cell temperature fixed at 300 K.

The maximum output power density increases with temperature for both emitters [Figure 4(b)] and is nearly the same for both. For example, the optimized emitter delivers 0.41 W cm$^{-2}$ at 1300 K and 1.95 W cm$^{-2}$ at 1600 K, compared with 0.41 at 1300 K and 2.03 W cm$^{-2}$ at 1600 K for SiC emitter. The power density is set by the in-band emission, for which the two emitters have similar hemispherical emissivities.

The spectral efficiency [Figure 4(c)] explains the efficiency gain. For the optimized emitter, the spectral efficiency is 59.1% at 1100 K and exceeds 69% at and above 1300 K, reaching 72.9% at 1500 K. For SiC, it is only 23.9% at 1100 K and 47.7% at 1500 K. The largest difference, 35.3%, occurs at 1100 K. Correspondingly, the fraction of the absorbed power that lies below the bandgap [Figure 4(d)] decreases from 0.71 for SiC to 0.29 for the optimized emitter at 1100 K, and from 0.38 to 0.09 at 1500 K. Such reduced parasitic heating below the bandgap is expected to greatly reduce the requirement on thermal management of the cell, leading to improved overall efficiency of the whole system.

## IV. DISCUSSION

Our results show that a highly selective emitter can keep sub-bandgap power out of a conventional cell as effectively as advanced cell architectures do[16]. Equation (1) shows why. Below the bandgap, where $\varepsilon_e$ and $\varepsilon_c$ are both small, $\varepsilon_{\text{eff}} \approx \frac{\varepsilon_e \varepsilon_c}{(\varepsilon_e+\varepsilon_c)}$, which is smaller than either of the two. When the emitter emits strongly below the bandgap, $\varepsilon_e \gg \varepsilon_c$, so $\varepsilon_{eff} \approx \varepsilon_c$, and the sub-bandgap loss is set by the cell alone. This is the case for SiC, whose sub-bandgap emissivity is about 0.71, compared with a cell absorptance of about 0.07. Reducing this loss then requires a more reflective cell, such as an air bridge cell[15,16]. With a selective emitter, $\varepsilon_e$ is small as well, so $\varepsilon_{eff}$ falls well below $\varepsilon_c$. For our cell, the Au rear mirror gives a sub-bandgap reflectance of 93%, with the selective emitter the fraction of sub-bandgap power kept out of the cell reaches 99%. A selective emitter therefore relaxes the requirements on the cell. A planar emitter can be made by standard thin-film deposition without photolithography [17,18]. This also shows that spectral control on the emitter side remains important for a reflective cell. With the same cell of 93% sub-bandgap reflectance, the selective emitter raises the efficiency at 1500 K from 30.8% to 47.0%, by keeping 99% sub-bandgap power out of the cell.

To put the performance in context, we also compare our emitter with an ideal step emitter, which has unity emissivity above the bandgap and zero emissivity below it, at all angles, facing the same cell. Under the same model, the ideal step emitter gives 50.0% efficiency at 1500 K, and a blackbody emitter gives 32.7%. The optimized emitter thus reaches 94% of the efficiency of the ideal step emitter. The remaining gap comes from the sub-bandgap resonances and from the in-band emissivity, which is below unity and falls at large angles. The ideal step emitter delivers 1.61 W $cm^{-2}$ at 1500 K, about 30% more than the optimized emitter, which shows that the in-band emissivity limits the power density rather than the efficiency.

Table I compares our result with representative TPV studies. The experimental records have been achieved with nearly gray emitters and advanced cells [12,14–16], while multilayer emitters with low-bandgap cells, including GaSb, have been predicted to reach 24–27% [27,28]. Higher values have been predicted for optimized thin-film cells facing a black emitter[40] and for radiative-limit, normal-incidence models[28]. Since the models differ in their assumptions, such as the view factor, the treatment of the cell and the angular integration, the comparison is only indicative. Nevertheless, it shows that a planar selective emitter combined with a conventional cell can, in principle, reach efficiencies comparable to those of air-bridge[15,16] and multijunction cells[14] at moderate emitter temperatures.

**TABLE I.** Comparison with representative TPV studies. Exp.: experiment; Calc.: calculation. $T_e$ is the emitter temperature.

| **Ref.** | **Emitter** | **Cell** | **$T_e$ (K)** | **$\eta$ (%)** | **Type** |
|---|---|---|---|---|---|
| [12] | Graphite | InGaAs (0.75 eV), Au mirror | 1480 | 29.1 | Exp. |
| [15] | SiC | InGaAs (0.74 eV), air bridge | 1455 | 32.0 | Exp. |
| [16] | SiC | InGaAsP (0.9 eV), air bridge | 1708 | 43.8 | Exp. |
| [13] | Graphite | GaInAs (0.75 eV), Au mirror | 2123 | 38.8 | Exp. |
| [14] | W filament | GaAs/GaInAs tandem (1.4/1.2 eV), Au mirror | 2673 | 41.1 | Exp. |
| [20] | Si photonic crystal on MgO | InGaAs (0.70 eV) | 1338 | 11.2 | Exp. |
| [21] | Pt/$Al_2O_3$ metamaterial + filter | InGaAs (0.6 eV) | 1328 | 24.1 | Exp. |
| [46] | Blackbody | InGaAs thin film, AR + spacer + Au | 1500 | 51.7 | Calc. |
| [28] | W/$HfB_2$–$HfO_2$ multilayer + filter | GaSb (0.72 eV) | 1473 | 27.4 | Calc. |
| [27] | $SiN_x$/$SiO_2$/$TiO_2$ on W | 0.55-0.72 eV cell | 1773 | 24 | Calc. |
| [30] | 7-layer $SiO_2$/Si on W | GaSb ideal cell (0.73 eV) | 1473 | 41.1 | Calc.[c] |
| This work | 16-layer MgO/Si on W | GaSb (0.725 eV), AR + Au | 1300 /1500 | 43.1 / 47.0 | Calc.[b] |

[b]Hemispherical, polarization-resolved multiple reflections, Auger recombination, unity view factor. [c]Shockley–Queisser analysis.

The assumptions used to evaluate selective emitters strongly affect predicted efficiency. The 58.9% predicted in Ref. [29] was obtained with normal-incidence emissivity and the radiative limit of the cell and the spectra are shown only up to 5 μm. At 1273 K, about 23% of the blackbody power lies beyond 5 μm, and any emission of the emitter in this range adds to the heat absorbed by the cell without generating current. Similarly, the spectral efficiency of the $SiO_2$/Si emitter in Ref. [30] was evaluated up to 8 μm, which excludes the phonon absorption band of $SiO_2$ near 8–10 μm, and its system efficiency was estimated for an ideal cell in the Shockley–Queisser limit, with radiative recombination only. In this work, we have included the full spectral range from 0.5 to 20 μm, hemispherical emission, both polarizations, the real cell with its finite OOB reflectance, and Auger recombination.

We note modeling assumptions may lead to an overestimate of efficiency. We assume unity view factor, neglect Shockley–Read–Hall recombination, series resistance, and use room-temperature optical constants. At high temperature, free-carrier absorption in Si[30] and the increased emissivity of W can raise the sub-bandgap emission of the emitter [25]. The thermal stability of the emitter also needs experimental study. The melting point of bulk Si is 1687 K [47], so the results at 1700 K mainly illustrate the trend, and practical operation of this material system is limited to lower temperatures. Multilayer emitters containing Si have shown degradation at 1300 K [26], and the thermal-expansion mismatch between MgO ($11.5 \times 10^{-6}$ $K^{-1}$) and W ($4.3 \times 10^{-6}$ $K^{-1}$) [18], as well as possible interdiffusion and silicide formation at the W/Si interface[48], may cause degradation during thermal cycling [17,37]. Thin oxide diffusion barriers and capping layers, such as $HfO_2$ or $Al_2O_3$, have been shown to improve the stability of refractory multilayers [17,18,37]. Such layers have little effect on the present design. Adding a 20 nm $HfO_2$ capping layer on top and a 20 nm $HfO_2$ layer between the last Si layer and W reduces the efficiency at 1500 K only from 47.0% to 46.1% and the output power density from 1.24 to 1.13 W $cm^{-2}$. All-oxide multilayers containing MgO have also shown structural changes after 1 h at 1500 °C [38], which indicates that long-term tests at the operating temperature are needed. We note that the efficiency gain over SiC is already 19.6% at 1300 K and 20.8% at 1200 K, so most of the benefit is retained at moderate temperatures where stability is less demanding.

## V. CONCLUSIONS

This study predicts that emitter-side spectral engineering can substantially reduce sub-bandgap absorption in a conventional GaSb TPV cell. A numerically optimized, 16-layer aperiodic MgO/Si multilayer on tungsten suppresses 97.1% of the sub-bandgap blackbody-equivalent power at 1500 K. With reflection from the Au-backed GaSb cell, only 1.0% of this power is absorbed parasitically in the cell, compared with 6.7% for the SiC-emitter reference. In the present model, which accounts for hemispherical, polarization-resolved radiative exchange and Auger recombination, the corresponding conversion efficiency is 47.0% at 1500 K (43.1% at 1300 K), while the 1500 K maximum power density remains close to that of the SiC reference (1.24 versus 1.27 W $cm^{-2}$). This result supports a complementary route to TPV spectral management through selective thermal emission rather than exclusively through high-reflectance cell architectures.

**ACKNOWLEDGMENTS**

L.Z. acknowledges support from the Institute of Energy and the Environment Seed Grant through the Energy 2100 initiative, and start-up funding from Pennsylvania State University. B.P. acknowledges support from the U.S. Army DEVCOM ARL Army Research Office award # W911NF-23-2-0229. Computations for this research were performed on The Pennsylvania State University's Institute for Computational and Data Sciences' Roar Collab supercomputer. The views and conclusions contained in this document are those of the authors and should not be interpreted as representing the official policies, either expressed or implied, of the U.S. Army or the U.S. Government.